\documentclass[5p]{elsarticle}

\pdfoutput=1 

\makeatletter
\def\ps@pprintTitle{%
 \let\@oddhead\@empty
 \let\@evenhead\@empty
\def\@oddfoot{}%
 \let\@evenfoot\@oddfoot}
\makeatother

\biboptions{sort&compress}

\usepackage{amsmath}
\usepackage{epsfig}
\usepackage{xspace}
\usepackage{amssymb}
\usepackage{array}
\usepackage[colorlinks=true,
pdfborder={0 0 0}] {hyperref}
\usepackage{color}    
\usepackage[labelfont=bf]{caption}

\definecolor{darkred}{rgb}{0.6471, 0.1098, 0.1882}
\def\beqn#1{\begin{equation}\label{eq:#1}}
\def\eeqn{\end{equation}}

\def\beqa#1{\begin{eqnarray}\label{eq:#1}}
\def\eeqa{\end{eqnarray}}

\def\beq{\begin{eqnarray}}
\def\eeq{\end{eqnarray}}
\def\bea{\begin{eqnarray*}}
\def\eea{\end{eqnarray*}}

\newcommand{\eqn}[1]{Eq.~(\ref{#1})}
\newcommand{\eqns}[2]{Eqs.~(\ref{#1})-(\ref{#2})}
\newcommand{\Eqn}[1]{Eq.~(\ref{#1})}

\newcommand{\order}[1]{\mathcal{O}(#1)}

\newcommand{\fig}[1]{Fig.~\ref{fig:#1}}
\newcommand{\Fig}[1]{Fig.~\ref{fig:#1}}

\newcommand{\Ref}{Ref.}
\newcommand{\Refs}{Refs.}

\newcommand{\gsim}{\gtrsim}
\newcommand{\lsim}{\lesssim}
\def\vev#1{\left\langle #1\right\rangle}

\newcommand{\gev}{\,\mathrm{GeV}\xspace}
   
\newcommand{\kev}{\,\mathrm{keV}\xspace}
   
\newcommand{\mev}{\,\mathrm{MeV}\xspace}
\newcommand{\tev}{\,\mathrm{TeV}\xspace}

\usepackage[normalem]{ulem}

\begin{document}


\title{Minimal Asymmetric Dark Matter}
%
%

\author{Sofiane M. Boucenna}
\ead{boucenna@lnf.infn.it}
\author{Martin B. Krauss}
\ead{martin.krauss@lnf.infn.it}
\author{Enrico Nardi}
\ead{enrico.nardi@lnf.infn.it}

\address{ INFN, Laboratori Nazionali di Frascati, C.P. 13, 100044 Frascati, Italy}
\date{\today}

\begin{abstract}
  In the early Universe, any particle carrying a conserved quantum
  number and in chemical equilibrium with the thermal bath will
  unavoidably inherit a particle-antiparticle asymmetry. A new
  particle of this type, if stable, would represent a candidate for
  asymmetric dark matter (DM) with an asymmetry directly related to
  the baryon asymmetry.  We study this possibility for a minimal DM
  sector constituted by just one (generic) $SU(2)_L$ multiplet $\chi$
  carrying hypercharge, assuming that at temperatures above the
  electroweak phase transition an effective operator enforces chemical
  equilibrium between $\chi$ and the Higgs boson.  We argue that
  limits from DM direct detection searches severely constrain this
  scenario, leaving as the only possibilities scalar or fermion
  multiplets with hypercharge $y = 1$, preferentially quintuplets or
  larger $SU(2)$ representations, and with a mass in the few TeV
  range.
\end{abstract}

\maketitle

\section{Introduction}
\label{sec:intro}

The existence of dark matter (DM) is a well established fact, 
confirmed by a plethora of observations including the most recent
cosmological surveys~\cite{Ade:2013zuv}. However, so far all evidences
for DM come solely from gravitational effects, and its nature remains
yet to be understood.  If DM is constituted by new fundamental
particles, the most compelling question is perhaps which other types
of interactions these particles can have with ordinary matter, which
could allow its `discovery' via non-gravitational effects.
But the little we know about DM brings about other puzzles, and one of
the most intriguing ones is \textit{why is the DM energy density so close
  to the energy density of baryons: $\Omega_{DM}/ \Omega_{B}\approx  5.5$}~\cite{Ade:2013zuv}?

In recent years, numerous models and constructions have been put
forward in the attempt to explain this puzzle.  Two main classes of
models have been studied in the literature so far: asymmetric DM (ADM)~\cite{Nussinov:1985xr,Barr:1990ca,Barr:1991qn,%
Kuzmin:1996he,Hooper:2004dc,Farrar:2004qy,Kitano:2004sv,Nardi:2008ix,%
Kaplan:2009ag} with all its variants, and WIMP-based schemes, 
as for example the ones proposed in~\cite{Cui:2011ab, Cui:2012jh}
(see~\cite{Boucenna:2013wba,Petraki:2013wwa,Zurek:2013wia,Davoudiasl:2012uw}
for recent reviews).  These constructions usually rely on new
symmetries (for instance, in order to transfer the asymmetry) and/or
extended hidden sectors. It should be remarked, however, that
symmetries can just explain why the {\it number densities} are
comparable: $n_{DM}/n_{B} \approx \order{1}$, while the numerical
coincidence is in the {\it energy densities:} $\rho_{DM}/\rho_B \equiv
(m_{DM} n_{DM})/(m_N n_B) \approx \order{1}$. In most cases a suitable
value for $m_{DM}$ is chosen in order to reproduce the observations,
which means that the coincidence is not really explained. Models in
which an explanation is provided for the ratio of energy densities do
exist, but often rely on unusual
scenarios~\cite{Bousso:2013rda,Froggatt:2014tua}.

In this paper we investigate whether it is possible to relate the
baryon and dark matter number densities using just the gauge
symmetries of the standard model (SM).  Our framework assumes a
minimal ADM (MADM) sector but is otherwise fairly
model-independent. We assume that at temperatures well above the
temperature $T_{EW}$ of the electroweak (EW) phase transition, a CP
asymmetry is generated in the thermal bath (the origin of this
asymmetry is not relevant for us). At sufficiently low temperatures
($T \lsim 10^6\gev$) the rates of all SM interactions become faster
than the Universe expansion rate, and chemical equilibrium is enforced
among all SM particle species, that are thus characterized by
numerically similar density asymmetries.  We introduce in this
scenario a new $SU(2)_L$ multiplet $\chi$ carrying hypercharge, whose
lightest (neutral) component is rendered stable by a matter parity.
An effective operator ensures that at $T \gsim T_{EW}$ $\chi$ is in
chemical equilibrium with the Higgs multiplet, and thus it inherits an
asymmetry which, after the symmetric component has annihilated away,
is at the origin of its present relic density.  We will show that
limits from DM searches via direct detection (DD) experiments,
together with the requirement that the effective interaction goes out
of equilibrium before hypercharge symmetry gets spontaneously broken,
render this scenario quite constrained.  We find that the only viable
MADM candidates are fermion or scalar multiplets with hypercharge 
$y=1$.\footnote{Subleading contributions from an asymmetric DM component
  to $\Omega_{DM}$ are however possible also in other cases.}  An
important difference with respect to other ADM scenarios is that in
our case, while the DM relic density is indeed inherited from an
initial asymmetry, DM is no more asymmetric in the present
cosmological era.  This is because when the Higgs field acquires a
vacuum expectation value (vev) and hypercharge symmetry gets broken,
the same operator responsible for the asymmetry transfer generates a
splitting between the two real degrees of freedom $\chi_{1,2}^0$ of
the neutral component of the complex multiplet. DM corresponds to the
lightest state $\chi_1^0$ (a real scalar or a Majorana fermion) which
can well undergo self annihilation and produce indirect
detection signals.  This is clearly different from the cases in which
the present-day DM population is still characterized
by an asymmetry in a conserved quantum number, and
no signal from DM  annihilation is  
expected.\footnote{Even in this case, if  the symmetric component
has been only partially annihilated away, indirect detection signals,
although accordingly suppressed,  might still be detectable,
see e.g.~\cite{Graesser:2011wi,Bell:2014xta}.}

\section{Minimal Asymmetric Dark Matter}
\label{sec:scenario}

The particle content of our DM scenario is that of the SM augmented
with an $SU(2)_L$ multiplet containing a neutral component which
accounts for the DM. In this respect it might resemble the minimal
DM (MDM) scenario proposed
in~\cite{Cirelli:2005uq,Cirelli:2009uv}. However, while MDM considers
self-conjugate multiplets with zero hypercharge, we require non-zero
hypercharge to ensure that the DM multiplets are not self-conjugate,
and can thus carry a particle-antiparticle asymmetry.  This implies
that the phenomenology of MADM is genuinely different from
that of MDM.

As usual, in order to enforce DM stability, we need to impose a parity
symmetry under which $\chi$ is the only odd field.  A second important
requirement is DM neutrality.  An $SU(2)_L$ multiplet $\chi$ of weak
isospin $t$ and hypercharge $y$ (without loss of generality we take
$y$ to be positive) can contain an electrically neutral component if
$t=y+k$, with $k$ a non-negative integer.\footnote{Electric charge is
  defined here as $Q=t_3 + y$ with $t_3$ the diagonal generator of
  weak isospin.}  For the minimal case $k =0$, for which the multiplet
has the lowest dimension, the electrically neutral component
corresponds to to the lowest weight $t_3=-y$, while for non-minimal
multiplets with $k > 0$ the lowest weights are negatively charged. In
all cases we need to ensure the neutral component remains the lightest
one within the multiplet.  A mass splitting between the charged and
neutral component of $\chi$ can be generated after EW symmetry
breaking by any type of $\chi$ couplings to the Higgs involving $\chi$
bilinears that are not by themselves invariant under $SU(2)_L$.  For
scalars there is always such a renormalizable operator:
\begin{equation}
\label{eq:T3}
{\mathcal{O}^{\vec t}}= \lambda_v 
\left(\chi^\dagger\, \vec{t}\, \chi\right)\,\left(\phi^\dagger
  \frac{\vec{\tau}}{2}\phi\right), 
\end{equation}
where $\vec{t}$ are the $SU(2)$ matrices in the representation in
which $\chi$ transforms and $\vec{\tau} $ are the Pauli matrices.
After EW symmetry breaking $\mathcal{O}^{\vec t}$ induces a mass
difference between two $\chi$ components of isospin eigenvalues $t_3$
and $t'_3$ given by:
\def\ltev{\lambda^{\scriptscriptstyle 1 {\rm TeV}}_{\scriptscriptstyle 0.02}}
\begin{equation}
\label{eq:chargedsplit}
\delta m^v = -(t_3-t'_3) \frac{ \lambda_v v^2}{4 m_\chi} \approx 
- 151 \; (t_3-t'_3)\, \ltev \; \mev,
\end{equation}
where
$ v=\langle \phi\rangle = (2\sqrt{2} G_F)^{-1/2} \approx 174 \gev$ is
the Higgs vev and we have defined
$\ltev = \frac{\lambda_v}{0.02}\, \frac{1\tev}{m_\chi}$.  For fermions
the operator corresponding to \eqn{eq:T3} is of dimension five, and
the result \eqn{eq:chargedsplit} holds with the replacement
$1/m_\chi \to 1/\Lambda$.  Neutral-charged mass splittings receive
also contributions from gauge boson loops. We obtain:
\begin{align}
\nonumber
\delta m^{\alpha_2} &= \frac{\alpha_2}{2}\left(t_3-t'_3\right)
\left\{\left(t_3+t'_3\right)\left({M_W}-c^2_W{M_Z}\right)
+2 y s_W^2 M_Z\right\} \\
\label{eq:Coulomb} 
&= 152 \; (t_3-t'_3)\,\left\{1.1(t_3+t'_3)+ 4.6 y\right\} \mev\,,
\end{align} 
with $\alpha_2=\frac{g^2}{4\pi}$ and ($c_W$) $s_W$ the (co)sine of the
weak mixing angle.  \Eqn{eq:Coulomb} agrees with the result given
in~\cite{Cirelli:2005uq} and holds for scalars as well.  We see that
for minimal multiplets (those with $t_3^\text{min} =-y$)
$\delta m^{\alpha_2}$ and (for $\lambda_v<0$) also $\delta m^v $ shift
the mass of the charged components above the mass of the neutral one
(e.g. for a scalar triplet with $y=1$  and 
reference values of the parameters, the mass 
splittings between the $Q=+1$ and $Q=0$ components are  
$\delta m^{\alpha_2} \sim 540\mev$ 
and,  for negative $\lambda_v$, $ \delta m^v \sim 151\mev$). 

For non minimal multiplets ($t_3^\text{min} = -(y+k)$) states of
weight $-(y+l)$ with $1\leq l \leq k$ are all (negatively) charged.
Among them, loop corrections would make heavier than the $t_3 =-y$
neutral state only those with $l > 2.3\, y$.  Since $y=1/2$ is the
minimum hypercharge value allowing for a neutral component in the
multiplet, and since by definition a charged state with $l=1$ is
always present, if loop induced mass splittings were dominant, all
non-minimal multiplets would remain excluded as DM candidates.
However, including the tree level contribution $\delta m^v$ allows to
evade this conclusion. We find that the neutral state is always the
lightest one for positive values of $\ltev$ falling within the
interval:
\begin{equation}
\label{eq:ltev}
\ltev = 2.5 y \pm 1.1 \,. 
\end{equation}

Let us now assume that a (non-Hermitian) effective operator of dimension
$d\geq 4$ mediates an interaction between a pair of $\chi$ particles
and the Higgs field $\phi$. Since the hypercharge of the Higgs is
$y(\phi)=-1/2$ this operator takes the form
\begin{equation}
\mathcal{O}^\phi = \frac{1}{\Lambda^{4y-x}}\; \chi \chi \phi^{4y} \,, 
 \label{eq:nonreno}
\end{equation}
\noindent 
where $x=1\,(2)$ if $\chi$ is a fermion (boson), $y = y(\chi)$ is the
hypercharge of the $\chi$ particle, and $\Lambda$ is the scale where
the effective operator is generated.\footnote{In \eqn{eq:nonreno} we
  have implicitly absorbed in the scale $\Lambda$ an overall coupling
  $\lambda$ multiplying the effective operator. Of course, any bound
  derived on $\Lambda$ should be then understood as a bound on
  $\Lambda/ \lambda^{1/(4y-x)}$.  For the fermion doublet case ($x=1$,
  $y=1/2$) this operator was already used in \cite{Servant:2013uwa} to
  relate DM and the baryon asymmetry.}
The operator $\mathcal{O}^\phi$ plays two roles:
\begin{itemize}
\item At $T > T_{EW}$: $\mathcal{O}^\phi$ can enforce chemical
  equilibrium between $\phi$ and $\chi$, communicating the asymmetry
  present in the thermal bath to the DM sector.
\item At $T< T_{EW}$: $\mathcal{O}^\phi$ generates a mass splitting
  between the two real degrees of freedom $\chi^0_{1,2}$ of the neutral
  component of the complex multiplet:
\begin{equation}
\delta m^x_0 = \frac{v^{4y}}{\Lambda^{4y-x}}
\label{eq:split}
\end{equation}
\noindent 
where for fermions ($x=1$) $\delta m_0 \equiv m_{\chi_2^0} -
m_{\chi_1^0}  $ while for bosons ($x=2$) $\delta m^2_0
\equiv m^2_{\chi_2^0} - m^2_{\chi_1^0}\approx 2 m_{\chi} \delta m_0$.
\end{itemize}
Let us comment on the previous two points.  For definiteness, in the
first point we have assumed that some baryogenesis mechanism generates
an asymmetry in the SM sector, which is then communicated to the
$\chi$ sector via the operator $\mathcal{O}^\phi$.  We stress however,
that the opposite possibility is also viable. The main difference
would simply be that the fundamental asymmetry is no more in the SM
$B-L$ charge, that remains exactly conserved and with vanishing
asymmetry, but in a global hypercharge asymmetry of the SM particles,
which is exactly compensated by an equal in size and opposite in sign
asymmetry in the $\chi$
sector~\cite{Antaramian:1993nt,Sierra:2013kba}.

Note that the requirement of gauge invariance allows to write other operators
suitable to enforce chemical equilibrium between $\chi$ and the SM
particles. Of course, operators of higher dimension are not relevant
and can be neglected, however, for integer $y$'s, the operator
\begin{equation}
\label{eq:eR}
\mathcal{O}^{e_R} = \frac{1}{\Lambda^{3y-x}} \chi\chi (e_R e_R)^y  \,, 
\end{equation}
where $e_R$ is any of the SM $SU(2)_L$ singlet leptons (with
$y(e_R) = -1$) is  allowed, and its dimension is $y$ units lower
than the dimension of $\mathcal{O}^\phi$ in \eqn{eq:nonreno}.
Motivated by minimality, one could assume that the ultra-violet
realization of the model is such that operators of this type are
either forbidden, or that they are suppressed by additional powers of
$\Lambda$ with respect to naive power counting.  However, for
completeness, in section \ref{sec:hypercharges} we will briefly
comment on the effects of $\mathcal{O}^{e_R}$  in the case of $y=1$
multiplets, which include the interesting cases of fermion and scalar
triplets.

The second role played by $\mathcal{O}^\phi$ after EW symmetry
breaking is also of fundamental importance: the lightest new particle
$\chi_1^0$ (a real scalar or a Majorana fermion) does not couple to
the $Z$ boson, but virtual $Z$  exchange can mediate the
inelastic transition $\chi_1^0 \to \chi_2^0$.  In order to evade the
stringent limits imposed by direct searches for DM scatterings off
nuclei, we need to ensure that in most cases the kinetic energy of the
incoming DM particle will not suffice to trigger the inelastic
scattering, so that the rate of events gets kinematically suppressed
below the observable level. This implies a lower limit on the mass
splitting:
\beqn{splitting}
\delta m_0 = 2 m_\chi \left(\frac{v}{\Lambda}\right)^{4y}\,
\left(\frac{\Lambda}{2 m_\chi}\right)^x\gtrsim \delta m^{\rm min}\,.
\eeqn
Values of $\delta m^{\rm min}$  have been derived in~\cite{Nagata:2014aoa} 
for different DM masses and different hypercharges $y$.
In the DM mass range relevant for us they can be roughly parameterized
as $\delta m^{\rm min} \sim (1+0.2 y)\times 175 \kev$ for
$m_\chi$ of order few TeV.

In order for DM to originate from the asymmetry present in the
primordial plasma, the following steps are required to occur in
sequential order of decreasing temperature:
\begin{enumerate}
\item At some temperature $T\gg T_{EW}$ the effective operator
  $\mathcal{O}^\phi$ mediates in-equilibrium reactions feeding an
  asymmetry between the SM sector and the $\chi$ sector.
\item At a certain temperature $T_a >  T_{EW}$ the rate of these
  reactions drops below the Hubble rate $H$, and the $\chi$ sector gets
  chemically decoupled from the thermal bath.  The relevant effective
  Lagrangian at $T_a$ is then characterized by a global $U(1)_\chi$
  symmetry corresponding to rephasing of the $\chi$ field.  The
  quantity $Y_{\Delta \chi} \equiv Y_{\chi}-Y_{\bar \chi}$ (where
  $Y_{\chi}= {n_{\chi}}/{s}$, with $s$ the entropy density) is
  associated to the $U(1)_\chi$ global charge, and it remains
  conserved.
\item The annihilation $\chi \bar \chi \rightarrow SM$ that proceeds,
  for example, via (unsuppressed) gauge interactions, continues to
  erase the symmetric DM component until it freezes out at a
  temperature $T_s < T_a$.  After $U(1)$ hypercharge symmetry is
  spontaneously broken at $T_{EW}$ no conserved charge remains
  associated with the $\chi$ neutral members.  To avoid that the
  surviving ADM component will restart annihilating away via
  e.g. $\chi_1^0\chi^0_1 \to W^+ W^-\, (ZZ)$ mediated by $t$-channel
  exchange of $\chi^\pm$ ($\chi^0_2$), we  need to require $T_s >
  T_{EW}$.  If at $T_s$ $Y_{\bar\chi} \ll Y_{\Delta \chi} \approx
  Y_{\chi}$, then the present DM relic abundance is dominated by the
  initial $\chi$ asymmetry.
\item At some temperature $T_d < T_{EW}$, which depends on $m_\chi$
  and on the charged/neutral mass splitting $\delta m_\chi$,
  $\chi^\pm$ will decay to the lighter neutral states. Later on (but
  still safely before Big Bang Nucleosynthesis), also
  $\chi_2^0 \to \chi_1^0$ decays occur. Eventually, at $T\ll T_d$ we
  will have $Y_{\chi_1^0} = Y_{\Delta \chi} $ and the present DM
  energy density then is given by
  $\rho_\text{DM} = s\,m_\chi Y_{\Delta \chi} $.

\end{enumerate}

\noindent
Let us note the following: ($i$) if the annihilation of the symmetric
part $\chi \bar \chi \to SM$ proceeds mainly via gauge interactions,
freeze out occurs around $T_s \sim m_\chi/25$. The requirement
$T_s > T_{EW}$ (point 3.) then implies
$m_\chi \gsim 25\, T_{EW}$.\footnote{For scalars, annihilation can
  also proceed via renormalizable operators like $\mathcal{O}^{\vec t}$ and 
  $\lambda_s (\chi^\dagger \chi) (\phi^\dagger \phi)$. For
  particularly large couplings $\lambda_{v,s} > 1$ they could be dominant 
and yield $T_s < m_\chi/25$.} Estimating precisely the
SM value of $T_{EW}$ is a hard task, and for relatively large Higgs
masses $> 100\gev$ only a few studies
exist~\cite{Shanahan:1998gj,Moore:2000mx,Burnier:2005hp}.  In
particular, for a Higgs mass $\sim 125\gev$,
\Ref~\cite{Burnier:2005hp} quotes $T_{EW}\sim 130\gev$. Due to the
large uncertainties involved in these estimates we will conservatively
impose the condition $T_{EW}> 100\gev$ which yields the lower limit
$m_\chi \gsim 2.5\tev$.  As regards the freeze out of the interactions
mediated by the effective operator $\mathcal{O}^\phi$, we will take it
to be $T_a \sim m_\chi/10 > T_{EW}$. This value results in a Boltzmann
suppression that yields a MADM relic asymmetry in the ballpark to
account for $\Omega_{DM}$.

\subsection{Constraints from chemical decoupling} 
\label{sec:constraints} 

We now discuss, in a general way, the conditions under which $\chi$
can provide a DM candidate with a relic density originating from the
same primordial asymmetry giving rise to the cosmological
matter/antimatter asymmetry.  

The operator $\mathcal{O}^\phi$ in \eqn{eq:nonreno} induces two types
of reactions which can maintain $\chi$ in chemical equilibrium with the
thermal bath: $s$-channel annihilation $\chi \chi \leftrightarrow
{\phi^{4y}}$, and inelastic $t$-channel scattering\footnote{We thank S. Tulin
  for pointing out to us the relevance of the $t$-channel reactions.}
$\chi \phi^* \leftrightarrow \chi^* \phi^{4y-1}$.  We define $T_a$ as
the temperature at which chemical equilibrium cannot be any longer
maintained, which happens when the rates for both these reactions
\begin{align}
\label{eq:Gchi}
\Gamma_{\chi\chi} &= n_\chi^0 \vev{\sigma |v|}_{\chi\chi}\,, \\
\label{eq:Gphi}
\Gamma_{\chi\phi} &= n_\phi^0 \vev{\sigma |v|}_{\chi\phi}\,, 
\end{align}
become slower than the  Hubble expansion rate:
\beqn{outofeq}
\Gamma_{\chi\chi},\;  
\Gamma_{\chi\phi} 
\lsim H (T_a)\,.
\eeqn
After decoupling, the relic abundance of DM remains approximately
fixed.  Chemical decoupling of  $\chi$  always occurs in
the non-relativistic limit $T_a < m_\chi$ while the requirement
$T_a > T_{EW}$ implies that a relativistic number density is the one
appropriate for the Higgs boson. Thus,  the appropriate 
equilibrium number densities for \eqns{eq:Gchi}{eq:Gphi} are:
\begin{align}
\label{eq:n0chi}
n^0_\chi &= g_\chi\, \left(\frac{m_\chi T}{2\pi}\right)^{3/2} e^{-m_\chi/T} \,, \\
\label{eq:n0phi}
n^0_\phi &= g_\phi\, \frac{\zeta(3) T^3}{\pi^2}\,, 
\end{align}
with $g_\chi$ and $g_\phi$ the respective numbers of degrees of freedom  
and $\zeta(3)\approx 1.2$.
The thermally averaged cross sections for the two processes can be
estimated as:
\begin{align}
\label{eq:sigmaVchi} 
\vev{\sigma |v|}_{\chi\chi} 
&\sim \eta_\text{PS}^{(n)}\, m_\chi^{-2}\, 
\left(\frac{m_\chi}{\Lambda}\right)^{2(4y-x)}\,, \\
\label{eq:sigmaVphi} 
\vev{\sigma |v|}_{\chi\phi} 
&\sim 
\vev{\sigma |v|}_{\chi\chi} \, 
\left(\frac{T}{m_\chi}\right)^{4(2y-1)}\,, 
\end{align}
where $\eta_\text{PS}^{(n)}$ is a $n=4y$ body phase space numerical
factor.  
The temperature dependent multiplicative factor for the
$t$-channel process \eqn{eq:sigmaVphi} arises 
because  while for $s$-channel annihilation the available phase space
for the final states is determined by $m_\chi$, for the $t$-channel
is determined by the $\phi^*$ momentum, which is of 
order $T$.  We can now check by direct comparison 
which process, for the different cases, is the relevant one to maintain
chemical equilibrium down to $T_a$. The condition
$\Gamma_{\chi\chi} > \Gamma_{\chi\phi}$ is satisfied when 
\begin{equation}
\label{eq:za}
\frac{z}{\log z} \lsim 4(2y-1)+\frac{3}{2}\,, 
\end{equation} 
where we have defined $z = m_\chi/T$.  For $y=1/2$ this inequality is
never satisfied, so that the relevant processes enforcing chemical
equilibrium are  the $t$-channel scatterings.  For
$y=1$ $\Gamma_{\chi\chi}$ dominates as long as $z \lsim 15$. Since,
as mentioned above, the correct DM relic density is obtained if
chemical decoupling occurs around $z_a\sim 10$, for  $y=1$ 
$s$-channel annihilation is the most relevant process.
Finally, for $y > 1$ the decoupling temperature is always determined
by $\Gamma_{\chi\chi}$, and $t$-channel scatterings can be safely
neglected.
We thus need to consider separately the case $y=1/2$ (scalar and
fermion DM doublets belong to this class) from the cases with
$y\geq 1$ (scalar and fermion triplets belong to this class).  Let us
start from the latter case.

To evaluate $\Gamma_{\chi\chi}$ let us first estimate the value of
$n_\chi^0$ in \eqn{eq:Gchi} which would yield a correct DM relic
density.  Before the EWPT chemical equilibrium between the Higgs and
the DM multiplet imposes the condition:
\beqn{chem}
 \frac{\Delta n_\chi}{n^0_\chi} = - 2 y \frac{\Delta n_\phi}{n^0_\phi} \,,
\eeqn
where $\Delta n_\chi = n_\chi - n_{\bar\chi}$ and
$\Delta n_\phi= n_\phi - n_{\bar\phi}$, and the minus sign follows
from requiring consistency between the hypercharge assignments
$y(\chi) >0$, $y(\phi) < 0$ and hypercharge conservation.  By
normalizing both   asymmetries to the entropy density,
\eqn{eq:chem} can be rewritten as:
\beqn{chemY} 
\frac{Y_{\Delta \chi}}{Y_{\Delta \phi}}
=-2 y \frac{n^0_\chi}{n^0_\phi}\,.
\eeqn
Assuming the SM content of relativistic particles, the Higgs asymmetry
is related to  $\Delta_{B-L}$  by $Y_{\Delta
  \phi}=-\frac{8}{79}Y_{\Delta_{B-L}}$~\cite{Nardi:2005hs}. We
further have $Y_{\Delta_{B-L}}= \frac{79}{28} Y_{\Delta
  B}$~\cite{Harvey:1990qw} so that:
\beqn{DnY}
\frac{Y_{\Delta \chi}}{Y_{\Delta \phi}}
= -\frac{7}{2}  \frac{Y_{\Delta \chi}}{Y_{\Delta B}}
=  -\frac{7}{2}\omega\frac{m_B}{m_\chi}  \,,
\eeqn
where we have defined $\omega \equiv
\frac{\Omega_{DM}}{\Omega_B}$ and $m_B\approx 1 \gev$ is the
nucleon mass. Putting together \eqn{eq:chemY} and \eqn{eq:DnY} we obtain:
\begin{equation}
\label{eq:n0chi2}
 n_\chi^0 = \frac{7\,\omega}{4 y}\frac{m_B}{m_\chi} n_\phi^0\,.
\end{equation}
By means of \eqn{eq:sigmaVchi} and \eqn{eq:n0chi2}
the condition $\Gamma_{\chi\chi}\lsim H(T_a)$ 
translates to:
\begin{align}
\label{eq:foa1}
\nonumber
m_\chi^{-2} \; z_a^{-1} \;
\left(\frac{m_\chi}{\Lambda}\right)^{2(4y-x)}
\!\!\! &\lsim \frac{4\pi^3}{21\zeta(3)}\sqrt{\frac{\pi g_*}{5}}\frac{y}{\omega
\eta^{(n)}_{PS}} \frac{1}{M_P m_B}\   \\
&= 6.1 \, \frac{y}{\eta^{(n)}_{PS}} \times 10^{-19}\, \gev^{-2}  \quad 
\end{align} 
where $M_P=1.2 \times 10^{19}\,$GeV is the Planck mass, and we have
used $H = \left(4\pi^3g_*/45\right)^{1/2} {T_a}^2/ M_P$ for the Hubble
parameter with $g_*= 106.75$ the number of relativistic degrees of
freedom, and $\omega \approx 5.5$ from cosmological
observations~\cite{Ade:2013zuv}.
In the numerical analysis we have adopted for the phase space factor
the parametrization
$\eta_\text{PS}^{(n)}=1/[4\pi(3^3\times2^{11}\times\pi^4)^{2y-1}]$
which reproduces correctly 3-body and 4-body phase space when particle
multiplicities and identical particle final states are accounted for.

For $y=1/2$ the  condition   $\Gamma_{\chi\phi}\lsim H(T_a)$ 
yields instead:
\begin{equation}
\label{eq:foaDoub}
  m_\chi^{-1} z_a^{-1}\,\left(\frac{m_\chi}{\Lambda}\right)^{2(2-x)} < 5.9 
\times 10^{-16} \gev^{-1}\,.
\end{equation}

 
\subsection{DM multiplets with different hypercharges}
\label{sec:hypercharges}

For each value of the hypercharge $y(\chi)$, 
\eqn{eq:splitting} and \eqn{eq:foa1} (or \eqn{eq:foaDoub} if $y=1/2$)
provide strong constraints on the viable parameter space. Another
constraint that we will use follows from the requirement that the
effective operator~\eqn{eq:nonreno} provides a consistent description
of the interaction enforcing chemical equilibrium, which requires
$m_\chi < \Lambda$. Let us now study a few cases. 

For a fermion multiplet ($x=1$) with hypercharge $y=1$ (the minimal
choice is a complex $SU(2)$ triplet) the two
constraints~\eqn{eq:splitting} and~\eqn{eq:foa1} yield:
\begin{align}
  \label{eq:Lambdaf}
\Lambda &\lsim
\left(\frac{v^4}{\delta m^{\rm min}}\right)^{1/3} \approx 17\, \tev\,,  \\
  \label{eq:mchif}
m_\chi &\approx 10\, \left(\frac{\Lambda}{17 \tev}\right)^{3/2} 
\left( \frac{z_a}{10}\right)^{1/4}\, \tev\,,
\end{align}
where in the first equation we have used
$\delta m^{\rm min} \approx 200\kev$.  With $z_a \sim 10$ and taking
into account the limit on the cutoff scale \eqn{eq:Lambdaf} we obtain
$m_\chi\lsim 10\tev$ which, for $z_s \sim 25$, is completely
compatible with the requirement $T_{s} \gsim T_{EW}$. Therefore a
complex $SU(2)_L$ fermion multiplet with $y=1$ can be a viable MADM
candidate.  The relatively low value of $\Lambda$ implies that
dimension five operators yield a rather large charged/neutral tree
level mass difference $\delta m^v \sim 1 \gev $
which dominates over the loop contributions, and is also much larger
than the splitting $\delta m^0$ between the two neutral states
$\chi_{1,2}$.  (In the presence of the transfer operator
$\mathcal{O}^{e_R}$ \eqn{eq:eR} a $y=1$ fermion multiplet would still
be a viable MADM candidate within the narrower window
$2.5\tev \lsim m_\chi \lsim 6.7\tev$.)

\begin{figure}[t!]
\centering
  \includegraphics[width=\linewidth]{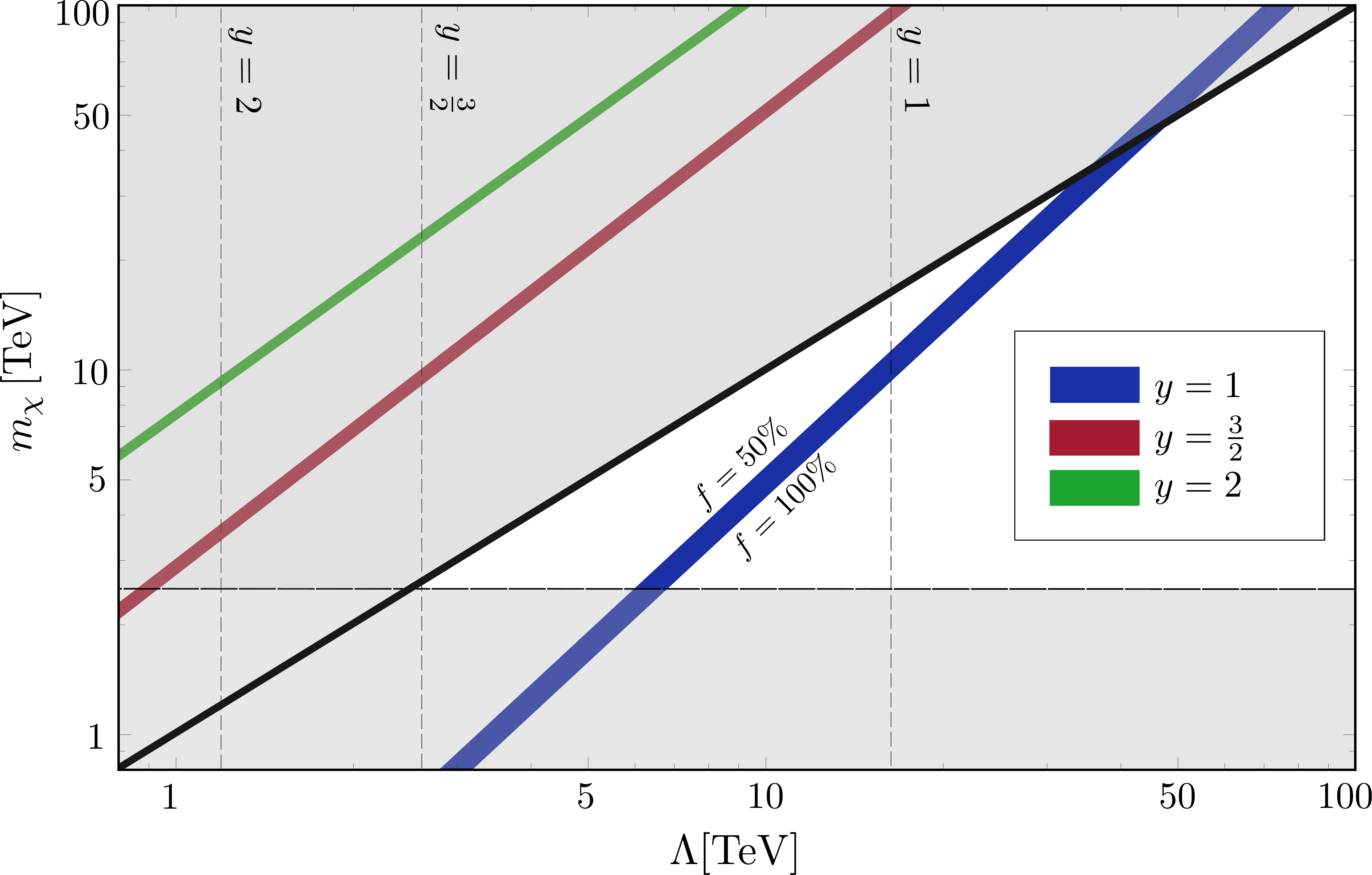}
  \caption{ Constraints on the parameter space $m_\chi$ \textit{vs.}
    $\Lambda$ for fermion DM with hypercharge $y = 1$ (blue),
    $\frac{3}{2}$ (red) and $2$ (green).  The width of the bands
    correspond to varying $f=\Omega_\chi/\Omega_{DM}$ in the interval
    $0.5< f<1.0$.  The region above the line bisecting the figure
    corresponds to $m_\chi > \Lambda$ for which the effective operator
    description \eqn{eq:nonreno} breaks down. This excludes the $y>1$
    lines.  The region below the dashed line is excluded by the
    requirement that all relevant reactions freeze-out above $T_{EW}$.
    The regions at the right of the vertical lines labeled
    $y=1,\frac{3}{2},2$ are excluded by DD
    experiments. \vspace{-0.2cm}}
\label{fig:modelspaceF}
\end{figure}

The results for fermions  in our MADM scenario for hypercharges
$y=1,\,\frac{3}{2},2$ are depicted in \fig{modelspaceF} (corresponding
to $\mathcal{O}^\phi$ operators respectively of dimension 7, 9, 
11).  The horizontal dashed line gives the lower limit on the
freeze-out temperature for $\bar \chi \chi$ annihilation
$T_s \sim m_\chi/25 > 100\gev$ and the gray region below is then
excluded. The thick black line bisecting the figure selects the region
$m_\chi > \Lambda$ (in gray) which must be excluded because the
description of the asymmetry transfer via the effective operator
$\mathcal{O}^\phi$ breaks down.  The regions on the right of the three
vertical lines corresponding respectively to $y=1,\,\frac{3}{2},2$,
delimit the values of $\Lambda$ that give a too large suppression of
the $\chi_2^0-\chi_1^0$ mass difference ($\delta m \lsim 200\kev$) so
that a signal would have been observed in  DD experiments.

The results for $\chi$ contributing dominantly to the DM of the
Universe are obtained from \eqn{eq:foa1}.  The width of the band
corresponds to varying the fraction of the relic abundance
$f\equiv\Omega_\chi/\Omega_{DM}$ from 50\% to 100\%.  As we have
discussed above, for a fermion multiplet with $y=1$ there is a region
up to $m_\chi \approx 10\tev$ and $\Lambda \approx 17 \tev$ in which
the $\chi$ relic density generated via an initial $\chi-\bar \chi$
asymmetry can account for the dominant amount of DM, while respecting
the other bounds.  The case $y=\frac{3}{2}$ corresponds to the red
band and $y=2$ to the green band.  In both cases the entire parameter
space selected lies in the $m_\chi>\Lambda$ half-plane, where the
effective field theory description of the asymmetry transfer cannot be
applied.
A first conclusion is that the case of a fermion multiplet with
$y = 1$ can be viable for a certain mass range, while for
hypercharges $y > 1$ the MADM scenario is not viable, or more
precisely the possible contribution of an asymmetry to the DM relic
density cannot be relevant.
%
For a fermion multiplet with hypercharge $y=1/2$ (the minimal
choice is an $SU(2)_L$ doublet) 
\eqn{eq:splitting} should be used together with~\eqn{eq:foaDoub}.  The
first condition yields the upper limit
$\Lambda \lsim 1.5 \times 10^5\tev$, while ~\eqn{eq:foaDoub} implies
the lower bound
$\Lambda \gsim 4.1\times 10^5 \left(\frac{T_a}
  {100\gev}\right)^{1/2}\tev$.
Given that $T_a > T_{EW} \gsim 100\gev$ the two bounds are in conflict, and 
we can conclude that for fermion doublets (and
more generically for fermion multiplets with hypercharge $y=1/2$) the
MADM scenario is not viable.

\begin{figure}[t!]
\centering
  \includegraphics[width=1.015\linewidth]{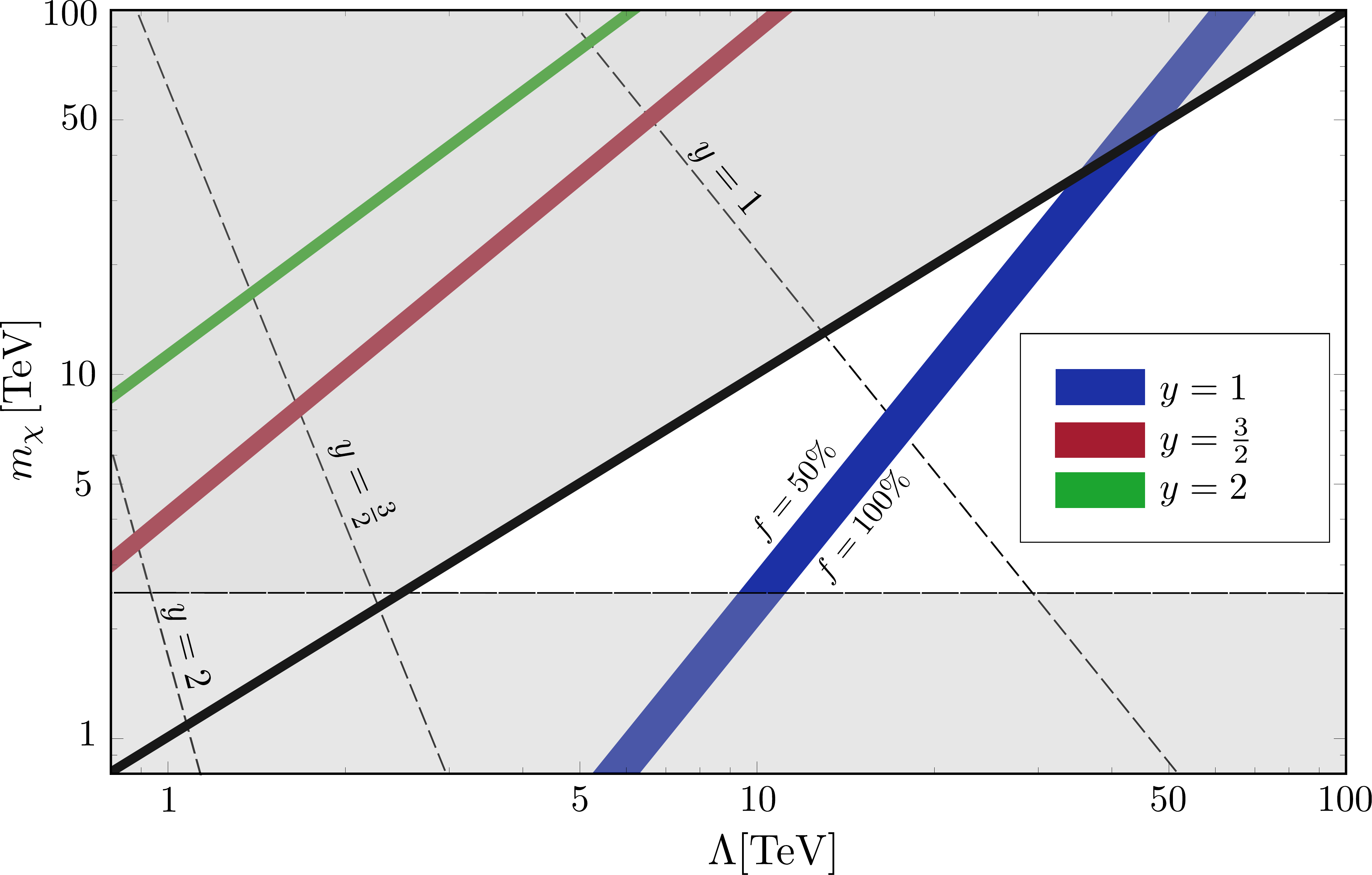}
\caption{Same as \fig{modelspaceF} for scalar DM.}
\label{fig:modelspaceS}
\end{figure}

For a scalar multiplet  ($x=2$) with hypercharge $y=1$ the two
constraints \eqn{eq:splitting} and \eqn{eq:foa1} yield:
\begin{align}
  \label{eq:Lambdas}
\Lambda &\lsim 
\frac{v^2}{\left(2 m_\chi \delta m^{\rm min}\right)^{1/2}} \,, 
\\
  \label{eq:mchis}
m_\chi &\approx \frac{\Lambda^2}{m_B} \left(6.1\times 10^{-14}\, z_a\right)^{1/2}\,.  
\end{align}
The value of $m_\chi$ is maximized by saturating the inequality 
\eqn{eq:Lambdas} in which case solving the system gives:
\begin{align}
  \label{eq:Lambdas2}
\Lambda &\approx 18 \left(\frac{10}{z_a}\right)^{1/8} \, \tev\,, 
\\
  \label{eq:mchis2}
m_\chi &\lesssim  6.7  \left(\frac{z_a}{10}\right)^{1/4} \, \tev\,.  
\end{align}
The last equation then allows for
$2.5\tev \lesssim m_\chi\lesssim 6.7\tev$ and shows that values of 
$m_\chi$ large enough to ensure that chemical equilibrium reactions
and annihilation processes freeze out before $T_{EW}$ are possible in
a rather large window.  (Asymmetry equilibration via the transfer
operator $\mathcal{O}^{e_R}$ \eqn{eq:eR} would instead imply
$m_\chi \lsim 1.3\tev$, and would render the $y=1$ scalar case not
viable.)
\Fig{modelspaceS} depicts the results for
scalar DM. Graphical conventions are the same as in \fig{modelspaceF}.
We see from the picture that the only case in which a DM asymmetry can
give sizable contributions to $\Omega_{DM}$ is for $y=1$.  For
higher values of the hypercharge the bands lie in the $m_\chi>\Lambda$
half-plane, and the corresponding MADM possibilities are therefore
ruled out.


For a scalar multiplet with hypercharge $y=1/2$ (e.g. a scalar
doublet), the operator $\mathcal{O}^\phi$ is of dimension four
(renormalizable). Thus there is no cutoff $\Lambda$ in the model and
$m_\chi$ is the only new scale, a feature that is unique to this
case. In order to pin down the values of $m_\chi$ it is
convenient to keep explicit the coupling constant $\lambda$ of the
transfer operator.  The constraint from DD, \eqn{eq:splitting},
implies the upper bound
$\frac{m_\chi}{\lambda} \lsim 8\times10^4\tev$, while the chemical
freeze-out condition~\eqn{eq:foaDoub} yields the lower limit
$ \frac{m_\chi}{\lambda} \gsim 4.1 \times 10^5
\,\left(\frac{T_a}{100\gev}\right)^{1/2}\tev$.\footnote{We
  thank the authors of~\cite{Dhen:2015wra} for helping us in spotting
  a numerical error in the phase space factor for this case.}
  The conflict between these
two bounds leaves no window in parameter space where scalar doublets
can work as MADM.

\section{Mass limits from symmetric annihilation}
\label{sec:symmetric}

We have seen in the previous sections that the bounds on the MADM
parameter space from (i) limits on nucleon recoils signals via
tree-level $Z$ boson exchange and (ii) constraints from the freeze-out
temperature of asymmetry transfer and annihilation processes, select
as the only possibilities multiplets with hypercharge $y=1$. The
minimal dimension of the corresponding representations are $SU(2)_L$
triplets, and the next-to-minimal are quintuplets.  An important issue
that should be discussed in more detail is which ranges of masses are
allowed by the requirement that $\chi \bar\chi$ annihilation will be
efficient enough to ensure that the contribution to $\Omega_{DM}$ of
any surviving symmetric component remains subdominant, i.e.
$\Omega_{\bar \chi} \ll \Omega_\chi \sim \Omega_{DM}$.  Estimating the
bounds on $m_\chi$ that follow from this argument is not a
straightforward task, since for $m_\chi\gg M_W$ the annihilation cross
section for $SU(2)_L$ multiplets is generically affected by
non-perturbative Sommerfeld enhancements, which can result in a
sizable suppression of the relic density.  One of the most studied
cases is that of an $SU(2)_L$ triplet with zero hypercharge, that is a
wino-like DM, $\widetilde W$.  With the tree level annihilation cross
section, $\Omega_{\widetilde W} = \Omega_{DM}$ is obtained for
$m_{\widetilde W} = 2.5\tev$~\cite{Cirelli:2005uq}. More refined
studies which include Sommerfeld and higher order corrections have
found sizable enhancements of the annihilation rate, so that the
condition $\Omega_{\widetilde W} = \Omega_{DM}$ is fulfilled for
larger values of the mass. For example,
\Refs~\cite{Hisano:2006nn,Cirelli:2007xd,Hryczuk:2010zi} quote mass
values in the range $2.7 \tev \lsim m_{\widetilde W} \lsim 3.0\tev$,
while more recent studies~\cite{Cohen:2013ama,Hryczuk:2014hpa} give
even higher values $m_{\widetilde W} \sim 3.1-3.2\tev$.\footnote{It is
  worth remarking at this point that for $\chi\chi$ annihilation into
  $(\phi\phi)^{2y}$
  the $U(1)_Y$ non-relativistic potential is repulsive, so that the
  rates for chemical equilibrating reactions will get suppressed
  rather than enhanced. This would raise the corresponding freeze-out
  temperature favouring the viability of the MADM scenario.}
For a fermion triplet with $y=1$ the tree level result quoted
in~\cite{Cirelli:2005uq} is $m_\chi \sim 1.9 \tev$, which is lower
than in the $y=0$ case because of the larger multiplicity of the
complex multiplet.  To our knowledge, no results have been reported in
the literature for a $y=1$ fermion triplet including Sommerfeld
enhancements, however we would expect even larger effects than in the
$y=0$ case.  This is because in the $T\gg T_{EW}$ limit the
interaction range of the Sommerfeld potential is determined by the
Debye screening length in the thermal plasma $\sim 1/(g_{1,2} T)$
(with $g_{1,2}$ the $U(1)_Y$ and $SU(2)_L$ couplings) rather than by
the inverse gauge boson mass $1/M_W$. Although for $y\neq 0$ one
expects that $SU(2)_L$ forces would result in non-perturbative
corrections similar to the $y=0$ case, the somewhat larger range of
$U(1)_Y$ interactions can further enhance the effect. All in all,
based on the results for the $y=0$ case we make the educated guess
that $\Omega_{DM}$ can be completely accounted for by a symmetric DM
component in the mass range $2.7\tev \lsim m_\chi \lsim 2.8\tev$.  To
the extent this is a reasonable estimate, then \fig{modelspaceF} shows
that not much space is left for relevant contributions from the
$\chi-\bar\chi$ asymmetry. For $y=1$ scalar triplets
  similar arguments can be put forth, except that the lowest order
  result $m_\chi \sim 1.6 \tev$ is a bit lower than in the fermion
  case, implying that the mass range in which an asymmetry could give
  relevant contributions to $\Omega_{DM}$  is accordingly
  reduced.


  It was found in the previous section that for $y=1/2$ multiplets the
  bounds from the two conditions \eqn{eq:splitting} and
  \eqn{eq:foaDoub} are in conflict: for fermions the the lower bound
  on the cutoff scale
  ($\Lambda \gsim 4.1\times 10^5 \left(\frac{T}{100\gev}\right)^{1/2}
  \tev$)
  is almost three times larger than the upper bound
  ($\Lambda \lsim 1.5\times 10^5\tev$).  For scalars the quantity
  $ \frac{m_\chi}{\lambda}$ is bounded by the same lower limit, which
  is about five time larger than the upper limit
  ($ \frac{m_\chi}{\lambda}\lsim  8\times 10^4\tev$).  The
  $y=1/2$ cases of lowest dimension are, however, of particular
  interest since they correspond to a fermion doublet (similar to a
  pure Higgsino) and to a scalar doublet (similar to the scalar DM
  candidate of the inert doublet model \cite{Ma:2006km}), and
  therefore it is worth checking if, in case the previous conflicts
  could be reconciled in some way, the $y=1/2$ doublet MADM scenarios
  could become viable.
  For fermion doublets a tree level estimate of the $\chi$ mass that
  could account for $\Omega_{DM}$ via freeze-out of symmetric
  annihilation yields $m_\chi\sim 1.2\tev$~\cite{Cirelli:2005uq}.
  Non-perturbative corrections to this result have been found to be
  negligible~\cite{Cirelli:2007xd}. Then the condition
  $m_\chi\gsim 2.5\tev$ that ensures that freeze-out of the relevant
  processes occur above $T_{EW}$ implies that the MADM relic density
  would largely overshoots the observed value of $\Omega_{DM}$.  For a
  scalar doublet the value hinted by symmetric annihilation
  $m_\chi\sim 0.54\tev$~\cite{Cirelli:2005uq} is also not affected
  much by Sommerfeld corrections,\footnote{Larger values of $m_\chi$
    are possible if annihilation into Higgs scalars largely dominates
    over annihilation mediated by gauge bosons.}  and the same
  conclusion holds.  All in all, the results of the previous section
  together with considerations of the $m_\chi$ values needed to
  realize the condition $\Omega_{\chi}\approx \Omega_{DM}$ via
  symmetric annihilation, indicate that the MADM scenario cannot be
  relevant for fermions or scalars with $y=1/2$.


  The general conclusion is that among multiplets of minimal
  dimension, only $y=1$ scalar/fermion triplets can marginally satisfy
  the condition of a sufficient suppression of the symmetric part of
  the relic density, so that the MADM scenario can become relevant.
  However, in the case of $y=1$ multiplets of higher dimension (e.g. a
  quintuplet) the MADM scenario, subject 
to the constraint \eqn{eq:ltev}, becomes more easily viable.  This is
  because the annihilation cross section for the symmetric component
  gets enhanced roughly as the fourth power of the multiplet
  dimension. This implies a strong suppression of the relic density,
  and correspondingly larger values of $m_\chi$ are required to
  saturate $\Omega_{DM}$ in the absence of an asymmetry.
%
%
  Moreover, for larger representations non perturbative corrections to
  the annihilation processes become particularly important. As an
  example, it was found in Ref.~\cite{Cirelli:2007xd} that for a
  fermion quintuplet with $y=0$, $m_\chi\sim 4.4\tev$ obtained at tree
  level~\cite{Cirelli:2005uq} gets boosted up to $m_\chi\sim 10\tev$
  after the inclusion of Sommerfeld effects~\cite{Cirelli:2007xd}.
  Therefore a thermally produced DM fermion quintuplet of mass
  $m_\chi \ll 10\tev$ could contribute the whole of DM only if its relic
  abundance is dominated by an initial asymmetry.

\section{Other phenomenological implications}
\label{sec:pheno}

Let us finally discuss briefly other possible phenomenological
implications of MADM candidates.

{\it Searches at colliders:} 
Searches at colliders of EW interacting new particles have been
performed, but the current reach of LHC is only of a few hundred
GeV~\cite{Chatrchyan:2012tea,%
  Chatrchyan:2012me, Aad:2013oja,Aad:2013yna,Aad:2014vka,%
  Khachatryan:2014mma} which does not constrain the interesting mass
range for MADM. Future $e^+e^-$ colliders with an energy reach of
$\sqrt{s}\sim 5\tev$ and $pp$ colliders with $\sqrt{s}\sim 100\tev$
will only marginally probe the multi TeV parameter
space~\cite{Chattopadhyay:2006xb,Low:2014cba,Cirelli:2014dsa}.  The
chances that MADM particles will be ever produced at foreseeable
colliders are thus rather feeble.

{\it Direct Detection experiments:}
While MADM tree level $Z$ mediated interactions with nuclei are
kinematically forbidden, non-vanishing DD cross sections appear at the
loop level. However, an accidental cancellation among various
contributions~\cite{Hisano:2011cs,Hill:2011be,Hill:2013hoa,Hill:2014yka}
results in suppressed cross sections $\sim
\mathcal{O}(10^{-47})\;$cm$^2$, which are by far below the current
experimental bounds~\cite{Aprile:2012nq,Akerib:2013tjd}.  In the
relevant mass range ($m_\chi \gsim \tev$), the cross sections remain
also below the reach of next generation DD
experiments~\cite{Cushman:2013zza} and close to the  neutrino
scattering background.

{\it Indirect detection:} 
The possibilities to bound (or discover) MADM via indirect
  detection (ID) of signals from DM annihilation are more optimistic.
  While it is well known that any conclusion derived from searches of
  DM annihilation byproducts heavily depends on the DM halo model,
  large portions of the mass range remain ruled out also when adopting
  rather implausible profiles.  The most relevant bounds come from
  cosmic-ray antiprotons measurements and from the absence of
  gamma-ray line features towards the galactic center.  For example,
  for a $y=0$ fermion triplet (wino-like DM) the corresponding bounds
  have been thoroughly studied,
  e.g. in~\cite{Cirelli:2007xd,Cohen:2013ama,Hryczuk:2014hpa},
  with the result that a mass range
  $1.8 \tev \lsim m_{\widetilde W}\lsim 3.5\tev$ is excluded.  There
  is a simple reason to expect that $y=1$ MADM triplets could be also
  strongly disfavoured by ID limits: the annihilation cross section
  gets a large enhancement from the formation of loose bound states
  when the range of the bounding interaction $\sim 1/M_W$ becomes of
  the order of the Bohr radius of the two particles state
  $\sim 1/(\alpha_2 m_\chi)$~\cite{Hryczuk:2010zi}, that is around 
  $m_\chi \sim M_W/\alpha_2 \sim 2.4\tev$, and we have seen
  that  $y=1$ triplets have the allowed values of $m_\chi$  rather close to
  this region.  The same conclusion does not apply, however, to
  non-minimal $y=1$ multiplets (e.g. quintuplets) for which the
  allowed mass range extends to $m_\chi \gg 2.5\tev$.  The issue of
  reliable MADM bounds from indirect detection for fermion and scalar
  quintuplets clearly deserves a specific study.  

\section{Conclusions}
\label{sec:conclusions}

Our study started from the observation that any new $SU(2)_L$
multiplet carrying non-vanishing hypercharge and in chemical
equilibrium with the thermal bath at $T>T_{EW}$, will be unavoidably
characterized by an asymmetry in its number density.  We have assumed
that some matter parity renders the lightest member of this
multiplet stable, thus providing a candidate for DM.  We have imposed a
requirement of {\it minimality}, that is that no other new particle is
introduced to help evading phenomenological constraints. We have also
explored under which conditions the present-day relic abundance of 
such a DM candidate can be (mostly) determined by its initial
asymmetry, which would justify denoting it as MADM.

A first set of constraints comes from limits from DD experiments,
which exclude DM candidates interacting via (unsuppressed) tree-level
$Z$ boson exchange. We have seen that this requirement can be
satisfied by our MADM candidates in a minimal way: a single effective
operator coupling DM to the Higgs field can in fact first be
responsible (at $T>T_{EW}$) of enforcing chemical equilibrium between
DM and the thermal bath, and next it can ensure that after EW symmetry
breaking the lightest mass eigenstate corresponds to a real scalar or
to a Majorana fermion, none of which couples (diagonally) to the $Z$
boson. Still, $Z$-mediated inelastic scatterings involving the
next-to-lightest neutral state impose severe constraints on viable
MADM scenarios. The requirement that reactions enforcing chemical
equilibrium, as well as DM annihilation processes decouple before the
EW phase transition, leaving the correct amount of DM, provides
another set of constraints. Together with the former ones, these allow
to exclude all MADM candidates except scalar and fermion multiplets
with hypercharge $y=1$, for which the lowest dimension representations
containing a neutral member are triplets.
However, even in this case, gauge annihilation could hardly erase
sufficiently the symmetric component, which will eventually constitute
most of the DM, and we have also seen that the mass range for which $y=1$
triplets could constitute good MADM candidates is already strongly
disfavoured by ID limits.  However, this conclusion does not
necessarily apply for larger representations.  For example, on the
basis of the analysis presented in~\cite{Cirelli:2007xd}, we have
argued that a (thermally produced) fermion quintuplet with $y=1$ and a
mass not much above the few TeV range, could account for the whole of
DM only if its relic number density is sufficiently enhanced by an
initial asymmetry.  This case is thus interesting, and we think it
deserves a dedicated study.

Finally, it should be mentioned that most of the constraints discussed
in this paper can be evaded by departing from minimality.  Perhaps the
simplest possibility is to add a SM singlet to which the `would be
MADM' can decay, thus transferring its asymmetry-related relic density
to a particle with no EW interactions.  This would automatically
bypass DD constraints and open up large portions of the parameter
space.  An ADM realization along this line involving an $SU(2)_L$
doublet fermion with $y=1/2$ decaying into a SM singlet has been put
forth for example in \Ref~\cite{Servant:2013uwa}.

\section*{Acknowledgments}
We acknowledge financial support from the research grant ``Theoretical
Astroparticle Physics'' number 2012CPPYP7 under the program PRIN 2012
funded by the Italian ``Ministero dell'Istruzione, Universit\'a e
della Ricerca'' (MIUR) and from the INFN ``Iniziativa Specifica''
Theoretical Astroparticle Physics (TAsP-LNF). S.M.B. acknowledges support
of the Spanish MICINN's Consolider-Ingenio 2010 Programme under grant
MultiDark CSD2009-00064.  E.N. acknowledges conversations with
R. Longas in the early stage of this work and 
A. Strumia for discussions. 


\providecommand{\href}[2]{#2}\begingroup\raggedright
 \endgroup

\end{document}